\documentclass[review]{elsarticle}

\usepackage{lineno,hyperref}
\modulolinenumbers[5]
\usepackage[margin=1in]{geometry}

\usepackage{graphicx}
\usepackage{listings}
\usepackage{makecell}
\usepackage[utf8]{inputenc}
\usepackage[T1]{fontenc}
\usepackage{multirow}
\usepackage{algorithm}

\usepackage{mathpazo} 
\usepackage{caption}
\usepackage{paralist}
\usepackage[justification=centering]{caption}
\usepackage{setspace}

\usepackage{graphicx} 
\usepackage{pgfplots}
\usepackage{pgfplotstable}
\usepgfplotslibrary{groupplots}
\usepackage{subcaption}

\pgfplotsset{compat=1.18}
\usepackage{amsmath}

\usepackage{algpseudocode}

\usepackage{sectsty}
\sectionfont{\fontsize{10}{10}\selectfont}

\journal{Computers \& Security}

\usepackage{amssymb}

\begin{document}

\begin{frontmatter}

\title{Explainable Machine Learning for Phishing Detection on Heterogeneous Datasets with MCP-Enabled Deployment
}

\author[1]{Nikhil Kumar Dora}
\ead{2481140@kiit.ac.in}

\author[1]{Sumit Kumar Tetarave}
\ead{sumitkumar.fca@kiit.ac.in}

\author[2]{Rishikesh Sahay}
\ead{rsaha@uis.edu}

\author[3]{Madhusudan Singh}
\ead{msingh@psu.edu}

\author[2]{Xiaoqing Li}
\ead{xli1@uis.edu}

\address[1]{School of Computer Applications,
Kalinga Institute of Industrial Technology, India}

\address[2]{Department of Management Information Systems, University of Illinois, Springfield, USA}

\address[3]{Department of Computer Science and Engineering, Pennsylvania State University, University Park, PA, USA}

\begin{abstract}

With the growth in digital transformation and Internet usage, the Social Engineering techniques such as Phishing have become a major concern for the users and the organizations.
Phishing attacks involve deceptive techniques to trick users into revealing confidential information that causes financial loss and reputation damage to organizations.
According to report of Verizon, 36\% of all data breaches involved phishing, highlighting the need for intelligent, adaptive, and explainable security mechanisms.
This paper examines the efficiency of different machine learning algorithms in phishing detection on heterogeneous phishing datasets that include a publicly available UCI dataset, our generated datasets using tools such as EvilGinx and Zphisher, and AI generated datasets.
Moreover, this work incorporates explainable AI (XAI) techniques such as Information Gain, SHAP (SHapley Additive Explanations), and LIME (Local Interpretable Model-Agnostic Explanations) to examine the most influential features impacting classification outcomes.
To support practical deployment, this work also incorporates an MCP-based phishing URL detection system that offers real-time URL analysis, feature extraction, confidence-based classification, and AI-assisted security interpretation. 
The experimental results demonstrate that among classical models the highest accuracy is obtained by Logistic Regression at 92.44\%, among ensemble models CatBoost achieved the highest accuracy at 95.01\%, among neural network CNN achieved an accuracy of 94.02\%, and among transformer-based models, DistilBERT got the highest accuracy at 99.78\%

\end{abstract}


\end{frontmatter}

\section{Introduction}
\label{sec:intro}

The advancements of the Internet technology and industry 4.0 offer great opportunities for individuals and businesses alike.
With the  growing trends of digital transformation in many industries, different cyber threats are also emerging.
Among the myriad tactics used by malicious actors, phishing attacks have emerged as a pervasive and insidious threat that exploits human vulnerabilities to compromise confidential information, financial assets, and critical infrastructure.
The Anti-Phishing Working Group (APWG) reported that in the second quarter of 2025, they recorded 1,130,393 phishing attacks~\cite{apwg}. 
This number increased from 1,003,924 attacks in the first quarter of 2025.
According to the CrowdStrike threat report, many cyber criminals used callback phishing to obtain initial access to the organizations in 2024~\cite{crowd_strike}.
The report also mentioned that cyber attackers are using Generative AI tools, particularly for social engineering attacks.
The CrowdStrike highlighted that LLM generated phishing messages have significantly higher chances (54\%) of getting clicked compared to messages written by humans.

Furthermore, Zscaler ThreatLabz reports a significant shift in the phishing landscape, i.e., from a high-volume spam campaign to sophisticated and targeted social engineering attacks, facilitated by Generative AI tools that can create highly realistic and convincing lures~\cite{zscalerthreatlab}.
The report further highlights that phishing campaigns use CAPTCHAs on malicious websites to evade automated security scanners and embed deceptive signals to bypass AI-powered detection.
This trend emphasizes the need for more resilient and explainable machine learning defenses.

Traditional phishing detection methods face limitations due to evolving attack techniques used by attackers.
Traditional detection methods depend on static features of webpages, which is insufficient to detect zero day phishing attacks~\cite{Zhang2025LeveragingML}. 
Therefore, advanced phishing detection techniques that can adapt to detect evolving sophisticated phishing attacks are required. 
Machine learning (ML) and Deep Learning (DL) offer this opportunity, making them capable of combating phishing attacks by efficiently identifying anomalies and patterns. 
Moreover, recent works have employed machine learning and deep learning for phishing detection by analyzing URL features of website~\cite{phishing_machine,backpropagation,phishing_url_detection,Zhang2025LeveragingML}.

Although previous studies have highlighted the effectiveness of machine learning and deep learning for phishing detection, many depend on single public datasets.
These may not capture the features of modern phishing URLs, including tool-generated phishing URLs and AI-generated URLs. 
To address this limitation, this study examines classical machine learning, ensemble learning, deep learning, and transformer-based models across multiple phishing datasets, while incorporating explainable AI techniques and an MCP-enabled phishing analysis system for practical deployment.

To capture diverse phishing patterns, four distinct data sources are utilized in this work: (a) a publicly available phishing dataset from the UCI Machine Learning Repository~\cite{uci_phishing}, (b) phishing URLs generated using open-source phishing tools such as EvilGinx, GoPhish, and Zphisher to simulate realistic phishing URLs, (c) phishing URLs downloaded from OpenPhish, and (d) synthetically generated phishing URLs with the help script provided by DeepSeek to analyze evolving phishing patterns. 
%
%
The main contributions of this paper are summarized below:

\begin{itemize}
\item Development of a comprehensive collection of phishing datasets.
\item Comparative performance analysis of multiple ML and DL models such as Random Forest, Decision Tree, AdaBoost, Logistic Regression, Multi-Layer Perceptron, and Deep Learning—across heterogeneous phishing datasets using standard evaluation metrics.
\item Application of SHAP, LIME and information gain based explainability to rank and analyze the most influential features impacting phishing detection across different datasets and learning models.
\item Design, implementation, and evaluation of a phishing MCP server using the best-performing ML model, assessed through advanced metrics such as Context Integrity Score (CIS), Attack Propagation Factor (APF), Mitigation Response Efficiency (MRF), and Context Sensitivity Index (CSI).
\end{itemize}

The rest of the paper is organized as follows:
Related extant literature on phishing detection is described in Section~\ref{sec:related_works}.
Section~\ref{methodology} describes the proposed methodology, including dataset generation, machine learning and deep learning models, and explainable AI metrics.
An implementation of phishing MCP server is explained in Section~\ref{mcp_implementation} along with its evaluation metrics.
Section~\ref{evaluation} presents the results.
Finally, Section~\ref{conclusion} presents the conclusion and future work.

\section{Related Works}
\label{sec:related_works}

Existing studies have widely explored the application of classical machine learning models using URLs, domain, and lexical features~\cite{phishing_domain, farea2025fsfs}.
The most widely used models are Random Forest, Decision Tree, Support Vector Machine, Naive Bayes, Logistic Regression, and AdaBoost. 
Most of these studies performed an evaluation on public datasets such as UCI, Kaggle, PhishTank, and Mendeley.
Several works using the UCI-based phishing dataset found that the Random Forest and Ensemble models are effective for phishing classification~\cite{phishing_domain,phishing_website}.
Other studies have also applied feature-selection-based Random Forest models, SVM-based classifiers, and proprietary machine learning approaches for phishing website detection~\cite{phishing_feature,lightweight_phishing,adaboost}.
Moreover, meta-learning and ensemble-based techniques such as AdaBoost-Extra Tree, Bagging-Extra Tree, Rotation Forest-Extra Tree, and LogitBoost-Extra Tree have shown improved detection accuracy of 97\% and low false-positive rates of 0.028~\cite{aimeta}.
However, many of these works depend on single public dataset that may reduce generalizability against evolving phishing attacks.

Recently, deep learning and transformer-based techniques have been studied for phishing and malicious URL detection. 
Deep learning models such as CNN, LSTM, and CNN-LSTM have been used to analyze URLs patterns, with CNN-based models often demonstrating better detection performance~\cite{deep_phishing}.
In~\cite{multidimensional}, a multidimensional deep learning technique combines URL character sequences, statistical URL features, webpage code features, and textual features to enhance phishing detection accuracy to 98.99\%~\cite{multidimensional}.
In~\cite{turk2025maliciousurl}, a transformer-based model is used for malicious URL detection, with fine-tuned ELECTRA model achieving almost perfect accuracy and outperforming many classical machine learning and deep learning baselines.
The authors pointed out the importance of lexical and structural URL features, show the superiority of transformer models for phishing and malicious URL detection, and highlight the need for future research on adversarial robustness and efficient real-time deployment.

Many recent works have tried to address these limitations by including expressive feature representations and conducting comprehensive comparisons across different models. 
For instance, PhishOFE integrates URL-based, HTML-based, and composite derived features and evaluates ten machine learning models, with CatBoost achieving an accuracy of 99.48\%~\cite{farea2025fsfs}. 
Similarly, in~\cite{nayak2025enhancing} the authors evaluated a large number of URL-based features and applied feature selection with deep learning architectures such as Feedforward Neural Networks (FNN), Deep Neural Networks (DNN) and TabNet~\cite{nayak2025enhancing}.
The work highlights the importance of robust feature engineering and model comparison. 
The results show that a feedforward model using only 14 selected features achieved an accuracy of 94.46\%.
Nevertheless, previous studies often focus on isolated benchmark datasets, limited dataset diversity, and insufficient explanation of model decisions.

In contrast, this work evaluates classical machine learning, ensemble learning, deep learning, and transformer-based models across heterogeneous phishing datasets, including publicly available datasets, tool generated phishing URLs, and AI-generated phishing URLs. 
Moreover, this work includes explainable AI techniques, including Information Gain, SHAP, and LIME, to study influential phishing features.
Furthermore, an MCP-enabled phishing analysis system is also developed to support practical deployment and real-time phishing URL analysis.

\section{Methodology}
\label{methodology}

\begin{figure}[h]
    \includegraphics[width=0.6\linewidth]{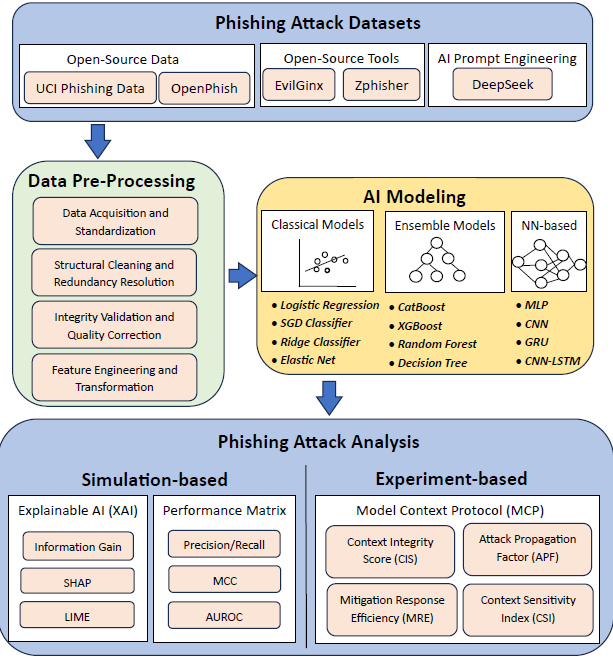}
    \centering
    \caption{Overall Framework}
    \label{proMethod}
\end{figure}

The proposed methodology uses multiple data sources to capture diverse patterns of phishing URLs and support robust phishing detection.
Figure~\ref{proMethod} shows the complete workflow, which includes data acquisition, preprocessing, feature standardization, model evaluation, and explainability analysis across machine learning, deep learning, and transformer-based models.
Each normalized dataset is evaluated using multiple machine learning and deep learning algorithms to classify and detect phishing attacks. 
Finally, the best performing model is evaluated using standard performance metrics and Model Context Protocol (MCP)-based analysis to test robustness, interpretability, and contextual reliability. 

\subsection{Phishing Datasets}
\label{sec:dataset}

\subsubsection{UCI Phishing Websites Dataset}

We used the publicly available UCI Phishing dataset as a baseline~\cite{uci_phishing}. 
It comprises 5,849 distinct samples (3,019 legitimate and 2,830 phishing) with 30 features that include basic URL and website attributes such as length, IP address presence, SSL status, ports, abnormal tags in HTML or Javascript, and redirect information.
Each instance is labeled as phishing (1) or legitimate (0).
This dataset served as a template for constructing features from other sources.

\subsubsection{Synthetic Phishing URL Generation Using EvilGinx and Zphisher}

To simulate phishing scenarios, we generate phishing URLs using tools such as \textbf{EvilGinx}~\cite{evilginx} and \textbf{Zphisher}~\cite{zphisher} within a controlled \texttt{Kali Linux} environment deployed in \texttt{VirtualBox}~\cite{sahay_phishing}. 
We provided a list of legitimate domains to these tools to generate the corresponding phishing links, and saved the phishing URLs and corresponding legitimate URLs in separate files. 
Links from these two separate files are combined into a single file, and features are extracted from the file, including phishing and legitimate URLs. 


\textit{EvilGinx} is used to generate legitimate looking phishing URLs that closely mimic real login pages of services like Google, Microsoft, and Facebook.
Furthermore, \texttt{Zphisher} is used to generate automated phishing URL, covering multiple social engineering templates, to introduce variations in the phishing dataset.
These tools enabled us to generate phishing URLs without interacting with live targets. 

Algorithm~\ref{alg:feature_based_generation} and Algorithm~\ref{alg:evilginx_automation} describe a detailed implementation of EvilGinx to generate phishing based dataset.  The dataset-generation pipeline uses two cooperating scripts: a Bash script show in Algorithm~\ref{alg:feature_based_generation} that generates feature-rich domain variants and an Expect-based EvilGinx automator, which is Algorithm~\ref{alg:evilginx_automation} that programmatically creates phishing lures and captures phishing URLs. 
The bash script calls the automator as a subprocess, collects phishing and legitimate URL pairs into output files, and performs post-hoc feature counting via regex-based analytics.
The workflow is executed in a virtual machine using authorized domains, with retry mechanisms to improve robustness.

\begin{algorithm}[t]
\caption{Feature-Based URL Generation using Bash Script}
\label{alg:feature_based_generation}
\begin{algorithmic}[1]
\Require Feature set $F = \{f_1, f_2, \ldots, f_k\}$, Target per feature $T$
\Ensure \texttt{phishing\_links.txt}, \texttt{legitimate\_links.txt}, \texttt{feature\_report.txt}
\State Clear previous output files
\State Create initial \texttt{feature\_rich\_domains.txt}
\For{each feature $f$ in $F$}
    \State Generate $T$ domain variants for $f$
    \State Append generated domains to \texttt{feature\_rich\_domains.txt}
\EndFor
\State Shuffle domain list for randomness
\For{each domain $d$ in domain list}
    \State success $\gets$ \textbf{False}
    \State retries $\gets 0$
    \While{(not success) and (retries < 3)}
        \State Run \texttt{evilginx\_automator.exp} with domain $d$
        \If{execution successful}
            \State success $\gets$ \textbf{True}
        \Else
            \State retries $\gets$ retries + 1
        \EndIf
    \EndWhile
\EndFor
\State Analyze \texttt{phishing\_links.txt} for feature presence
\For{each feature $f$ in $F$}
    \State Count occurrences of pattern corresponding to $f$
\EndFor
\State Write summary statistics to \texttt{feature\_report.txt}
\end{algorithmic}
\end{algorithm}

\begin{algorithm}[t]
\caption{Automated Phishing and Legitimate URL Collection using EvilGinx}
\label{alg:evilginx_automation}
\begin{algorithmic}[1]
\Require Number of links $N$, Domain base $D$
\Ensure Appends phishing URLs and legitimate URLs to output files
\State Launch EvilGinx in developer mode
\State Configure domain $\gets D$, IPv4 $\gets 127.0.0.1$
\State Enable (\texttt{phishlet})
\State Delete existing lures
\State Initialize output files for collecting phishing and legitimate URLs
\For{$i \gets 1$ to $N$}
    \State retries $\gets 0$
    \State url\_captured $\gets$ \textbf{False}
    \While{(not url\_captured) and (retries < 3)}
        \State Create new lure with EvilGinx
        \State Extract phishing URL from lure
        \If{URL is valid and not filtered}
            \State Generate matching legitimate URL for $D$
            \State Save both URLs to respective files
            \State url\_captured $\gets$ \textbf{True}
        \Else
            \State retries $\gets$ retries + 1
        \EndIf
    \EndWhile
\EndFor
\State Close EvilGinx session and output files
\State Return total successful URLs
\end{algorithmic}
\end{algorithm}

Similarly, Zphisher is used to create phishing URLs by selecting a set of pre-designed phishing templates and generate masked URLs in a controlled environment. 
The generated URLs are checked, stored, and matched with the corresponding legitimate URLs to create the dataset.
All activities are performed in an isolated virtual machine.

\subsubsection{AI Assisted Synthetic URL Generation using DeepSeek}

\begin{algorithm}[h!]
\caption{AI-Assisted Synthetic Phishing URL Generation}
\label{alg:phish_synthesis_nofeatures}
\begin{algorithmic}[1]
\Require Set of legitimate URLs $L$, target URL count $N$
\Ensure List of unique AI-assisted synthetic phishing URLs $U$

\State Initialize an empty list $U$ and hash set $H$
\State Define phishing URL transformation rules based on common malicious patterns

\While{$|U| < N$}
    \State Select a random legitimate URL $b$ from $L$
    \State Generate a phishing-like variant $v$ using one or more transformation rules
    \State Normalize $v$ and compute its hash value
    \If{hash of $v$ is not present in $H$}
        \State Append $v$ to $U$
        \State Insert hash of $v$ into $H$
    \EndIf
\EndWhile

\State Return synthesized URL list $U$
\end{algorithmic}
\end{algorithm}

DeepSeek is employed to support script generation to synthesize new phishing-like URLs in Google Colab to incorporate novel phishing patterns~\cite{deepseek_r1_2025}.
The generated URLs mimic common phishing traits such as subdomains with misleading names, words that suggest a security concern, and URL construction that is not obviously suspicious, as listed in Algorithm~\ref{alg:phish_synthesis_nofeatures}.

\subsection{Threat-Intelligence Based Phishing URL Collection}








Phishing URLs are also collected from threat-intelligence feeds such as OpenPhish and URLhaus, then deduplicated and processed for feature extraction.
All datasets are standardized using consistent feature names and labels, enabling reproducible experiments and straightforward comparison of various machine learning and deep learning approaches.
The overall class distribution in all the datasets are summarized in Table~\ref{tab:class_distribution}.


%

\begin{table}[ht]
\centering
\caption{Class Distribution for Each Dataset}
\label{tab:class_distribution}
\begin{tabular}{|l|l|l|}
\hline
\textbf{Dataset} & \textbf{Label (0: Legitimate)} & \textbf{Label (1: Phishing)} \\
\hline
UCI & 3019 & 2830 \\
OpenPhish & 1811 & 1884 \\
EvilGinx & 2446 & 2010 \\
GenAI & 2446 & 3200 \\
\hline
\end{tabular}
\end{table}

\subsection{Data Preprocessing}
\label{sec:Preprocessing}
We processed the data to ensure quality, consistency, and comparability across data sources. 
The dataset initially differed in structure, feature composition, and labeling format, required a unified preprocessing pipeline.

\subsubsection{Data Cleaning and Standardization}

In the preprocessing phase, an exploratory analysis of each dataset is performed, including examination of the shape, feature names, and label distribution of the dataset.
Initially, the UCI dataset contained 11,055 instances with 31 features showing a significant number of duplicate records. 
After removal of duplicates, the sample contains 5849 unique samples.
However, the OpenPhish, EvilGinx, and AI datasets do not contain duplicate records.
Moreover, the datasets are examined for missing or null values to ensure complete feature sets.

Furthermore, to ensure label standardization, in the UCI dataset, the target column \texttt{Result} with values $-1$ and $1$ is renamed as \texttt{label} for consistency. 
The remaining datasets already contain a binary \texttt{label} column ($0$ for legitimate and $1$ for phishing instances).

\subsubsection{Feature Alignment}

Feature inconsistency across datasets is addressed by standardizing feature names such as \texttt{URL\_Length} vs. \texttt{url\_length} and \texttt{popUpWindow} vs. \texttt{popup\_window}.
Additionally, redundant features are removed.
For example, in the EvilGinx dataset, duplicate representations of \texttt{URL\_Length} are resolved by keeping the numerical version and discarding the categorical version.
Initially, the datasets contain a varying number of features: 31, 37, 38, and 37 respectively. 
For consistency, a common feature space is established across all datasets.

\subsubsection{Coefficient-Based Feature Selection}

Feature selection is performed using a coefficient-based importance approach derived from Logistic Regression. 
This method is selected because of its computational efficiency, interpretability, and ability to provide a consistent global ranking of features. 
Given a feature vector $X = (x_1, x_2, \dots, x_n)$, Logistic Regression models the probability of phishing as:

\[
P(y=1|X) = \frac{1}{1 + e^{-(w^T X + b)}}
\]

where $w = (w_1, w_2, \dots, w_n)$ represents the learned coefficients.
The magnitude of each coefficient $w_i$ shows the importance of the corresponding feature.

To improve robustness, coefficient values are aggregated across stratified K-fold cross-validation as shown:

\[
\bar{w}_j = \frac{1}{k} \sum_{i=1}^{k} w_j^{(i)}
\]

where $k$ indicates the number of folds and $w_j^{(i)}$ shows the coefficient of feature $j$ in the $i$-th fold.

Depending on the averaged coefficients, a subset of 23 common and significant features is selected in all datasets.
It ensures uniformity in the input space and allowed fair comparisons between models.
Although coefficient-based methods mainly measure linear relationships, they can be used as an effective model-agnostic dimensionality reduction filter.

\subsubsection{Final Dataset Preparation}

In the final preprocessing stage, all datasets are reduced to 23 selected common features along with the standardized label column.
To ensure consistency and avoid bias, only the selected numerical features are used for model training.
The final preprocessed datasets offer a unified and reliable input structure for subsequent machine learning and deep learning models, enabling consistent evaluation and comparison across datasets.

\subsection{Machine Learning Algorithms}
\label{sec:ml_algo}

We use a diverse set of machine learning and deep learning algorithms to evaluate phishing detection on heterogeneous datasets.
In general, each model learns a decision function $f(\mathbf{x}) : \mathbb{R}^n \rightarrow \{0,1\}$ that maps an input feature vector $\mathbf{x}$ to a legitimate or phishing class. 
Model training aims to minimize the empirical risk 
$\frac{1}{N}\sum_{i=1}^{N}\ell(y_i, f(\mathbf{x}_i))$ given a labeled training set $\{(\mathbf{x}_i, y_i)\}_{i=1}^{N}$, where $\ell(\cdot)$ denotes the task-specific loss function. 
The formulation enables a fair and consistent comparison between different learning paradigms under identical experimental conditions.

\subsubsection{Regression-based Machine Learning Models}

Traditional linear and regression-based models are employed as baseline models because of their computational efficiency, interpretability, and good performance in high-dimensional feature space. 
Models evaluated in this category include: Logistic Regression, Ridge Classifier, Stochastic Gradient Descent (SGD) Classifier, Elastic Net–regularized Logistic Regression, and Linear Support Vector Classifier (Linear SVC). 
These models approximate the posterior probability $P(y=1 \mid \mathbf{x})$ using linear decision boundaries of the form 
$f(\mathbf{x}) = \sigma(\mathbf{w}^\top \mathbf{x} + b)$ or margin-based separation functions, where $\mathbf{w}$ shows the learned weight vector.

Feature standardization is applied to ensure stable optimization and comparable feature scales, specifically for gradient-based solvers. 
Regularization terms such as $\|\mathbf{w}\|_1$, $\|\mathbf{w}\|_2$, or their combination in Elastic Net are included into the loss function to reduce overfitting and handle correlated phishing indicators. 
Despite their simplicity, these models offer a good and interpretable baseline to quantify the performance gains introduced by MCP-driven context isolation and feature fusion.

\subsubsection{Tree-Based and Ensemble Models}

Tree-based and ensemble models are evaluated because of their capability to model complex, non-linear interactions among phishing related features. 
Unlike linear classifiers, these models partition the feature space using a hierarchy of decision rules, implicitly learning piecewise non-linear decision functions: 
$f(\mathbf{x}) = \sum_{m=1}^{M} \alpha_m h_m(\mathbf{x})$, Where $h_m(\cdot)$ indicates an individual decision tree and $\alpha_m$ its corresponding weight.

The models evaluated in this category include Decision Tree, Random Forest, Extra Trees, Gradient Boosting, XGBoost, and CatBoost. 
Bagging based ensemble methods such as Extra Trees or Random Forest, combine multiple trees that are de-correlated to reduce the variance of prediction, whereas boosting based methods iteratively minimize residual errors by focusing on misclassified samples.
CatBoost is included for its robustness to heterogeneous feature distributions and its ability to mitigate overfitting under distribution shifts,  which is particularly relevant for phishing datasets collected from real-world, tool-generated, and AI-synthesized sources.

\subsubsection{Deep Learning Models}

We explore multiple deep learning architectures, such as: Multilayer Perceptron (MLP), Convolutional Neural Network (CNN), Long Short-Term Memory (LSTM), Gated Recurrent Unit (GRU), and Bidirectional LSTM (BiLSTM) to capture higher-order feature interactions and latent contextual patterns. 
Deep neural networks learn hierarchical representations through successive non-linear transformations 
$\mathbf{h}^{(l)} = \phi(\mathbf{W}^{(l)}\mathbf{h}^{(l-1)} + \mathbf{b}^{(l)})$, enabling the detection of subtle phishing indicators that may not be linearly identifiable.

CNN-based models are used to learn localized and position-invariant feature patterns, while recurrent architectures, such as LSTM and GRU model sequential dependencies within the feature representations. 
BiLSTM further improves contextual understanding by processing feature sequences in both forward and backward directions. 
All deep learning models are optimized using the Adam optimizer and binary cross-entropy loss: 
$\ell(y,\hat{y}) = -[y\log(\hat{y}) + (1-y)\log(1-\hat{y})]$, with early stopping applied to prevent overfitting. 
This integrated mechanism ensures computational efficiency while maintaining robust generalization in various scenarios of phishing attack.

\subsection{ML Evaluation Metrics}

To improve model transparency and identify the most important phishing features, this study uses both statistical feature selection, including Accuracy, Precision, Recall, F1-Score, area under curve (AUC), Receiver Operating Characteristic (ROC)  and model-agnostic explainability techniques (XAI), including Information Gain, SHAP, and LIME.

\subsubsection{Confusion Matrix}

The confusion matrix for binary phishing classification is defined as:

\begin{equation}
\mathbf{M} =
\begin{bmatrix}
TP & FP \\
FN & TN
\end{bmatrix}
\end{equation}

where $TP$ indicates true positives (phishing correctly detected), $FP$ false positives (legitimate samples misclassified as phishing), $FN$ false negatives (phishing samples misclassified as legitimate), and $TN$ true negatives (legitimate samples correctly classified). 

Accuracy evaluates the overall correctness of the classifier and is defined as: $ (TP + TN)/(TP + TN + FP + FN)$. 
Precision quantifies the reliability of phishing predictions and is calculated as: $TP/(TP + FP)$.
Recall, referred to as the True Positive Rate (TPR), measures the classifier's ability to correctly identify phishing instances, i.e., $TP/(TP + FN)$. 
Moreover, the F1-score offers a harmonic mean of Precision and Recall, denoted by: $2 \times [(Precision \times Recall)/(Precision + Recall)]$.

\subsubsection{ROC and AUC}

The Receiver Operating Characteristic (ROC) curve demonstrates the trade-off between the True Positive Rate (TPR) and the False Positive Rate (FPR) across varying classification thresholds. The FPR is defined as: $FP/(FP + TN)$, and the ROC curve as: $\{(\text{FPR}(t), \text{TPR}(t)) \mid t \in [0,1]\}$, where $t$ denotes the decision threshold.
Moreover, the Area Under the ROC Curve (AUC) illustrates the discriminatory power of classifier and is defined as: $ \int_{0}^{1} \text{TPR}(\text{FPR}) \, d(\text{FPR})$. 
A higher AUC value denotes superior classification performance and robustness against threshold variation.

\subsubsection{Cross-Validation Evaluation}

To ensure statistical reliability, $k$-fold cross-validation is used. 
Let $S_k$ denote the performance score achieved from the $k$-th fold; the average performance metric is computed as: $\frac{1}{K} \sum_{k=1}^{K} S_k$.
This evaluation ensures an unbiased performance estimate across heterogeneous phishing datasets.

\subsubsection{Information Gain}

Information Gain (IG) evaluates the reduction in uncertainty of the target variable after observing a feature. 
Let $Y$ denote the class label (phishing or legitimate), and $X_j$ denote a feature. 
The entropy of $Y$ is defined as:

\begin{equation}
H(Y) = - \sum_{y \in \{0,1\}} P(y)\log_2 P(y)
\end{equation}

The conditional entropy of $Y$ given $X_j$ is:

\begin{equation}
H(Y \mid X_j) = \sum_{x \in X_j} P(x) H(Y \mid X_j = x)
\end{equation}

The information gain of the feature $X_j$ is then computed as:

\begin{equation}
IG(Y, X_j) = H(Y) - H(Y \mid X_j)
\end{equation}

Features with higher Information Gain contribute more significantly to reducing classification uncertainty and are prioritized during feature selection.

\subsubsection{SHAP (SHapley Additive Explanations)}

SHAP is a game-theoretic approach that assigns an importance value to each feature by fairly distributing the prediction outcome among all features. Given a model $f$ and an input instance $\mathbf{x}$, the SHAP explanation model is defined as:

\begin{equation}
f(\mathbf{x}) = \phi_0 + \sum_{j=1}^{n} \phi_j
\end{equation}

where $\phi_0$ represents the base value (expected model output), and $\phi_j$ denotes the Shapley value of feature $j$. The Shapley value is computed as:

\begin{equation}
\phi_j = \sum_{S \subseteq F \setminus \{j\}} \frac{|S|!(|F|-|S|-1)!}{|F|!}
\left[ f(S \cup \{j\}) - f(S) \right]
\end{equation}

where $F$ is the set of all features and $S$ is a subset that excludes the feature $j$. 
SHAP provides both global and local interpretability by quantifying the contribution of each feature to the final prediction.

\subsubsection{LIME (Local Interpretable Model-Agnostic Explanations)}

LIME explains individual predictions by approximating the original model locally with an interpretable surrogate model. For a given instance $\mathbf{x}$, LIME minimizes the following objective function:

\begin{equation}
\mathcal{L}(f, g, \pi_{\mathbf{x}}) =
\sum_{z \in Z} \pi_{\mathbf{x}}(z)
\left( f(z) - g(z) \right)^2 + \Omega(g)
\end{equation}

where $f$ is the original complex model, $g$ is the interpretable local surrogate model, $\pi_{\mathbf{x}}(z)$ is a proximity measure between instance $z$ and $\mathbf{x}$, $\Omega(g)$ penalizes model complexity.
LIME thus provides localized explanations by highlighting the most influential features responsible for individual phishing predictions.


\section{The Proposed Phishing MCP Server}
\label{mcp_implementation}

The Phishing MCP Server provides a middleware layer between phishing detection models and user-facing interfaces. 
The system is designed for modularity, scalability, and educational reproducibility, allowing researchers and students to interact with phishing-detection tools using simple API calls or web interfaces. 
The system is lightweight and requires minimal setup, making it practical for integration into research and LLM-based phishing-analysis workflows. 

\begin{figure}[h]
    \includegraphics[width=0.8\textwidth]{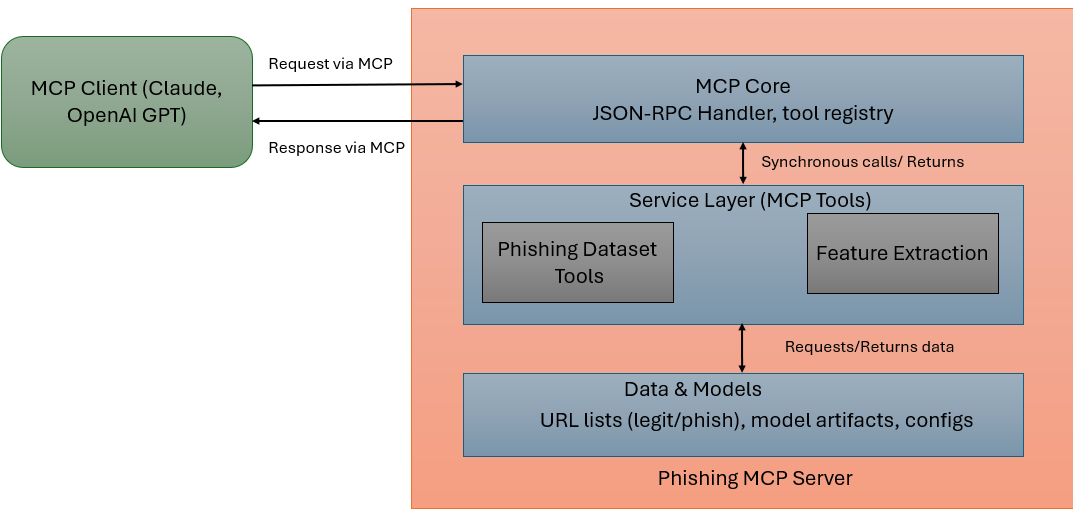}
    \centering
    \caption{Phishing MCP Server and its Integration with Client}
    \label{doc:phishing_mcp_server}
\end{figure}

Figure~\ref{doc:phishing_mcp_server} shows the proposed architecture with four conceptual layers.
 Client Layer encompasses any client that is compatible with MCP, such as ChatGPT, Claude Desktop, MCP Inspector, or web dashboards, which sends structured requests to the server for phishing analysis. 
MCP Core Layer  manages the standardized exchange of requests and responses via the MCP protocol. 
It handles tool registration, session management, and JSON serialization of requests and responses. 
Service Layer provides the core functionalities of the phishing detection system as MCP tools such as URL feature extraction and phishing model inference. 
Every tool executes its specific function and provides results in a structured JSON format, facilitating straightforward composition and integration. 
Data and Model Layer is the fourth layer, which comprises of datasets (legitimate and phishing URLs), and pre-trained machine learning models for real-time classification.



\begin{figure}[h]
    \includegraphics[width=0.7\textwidth]{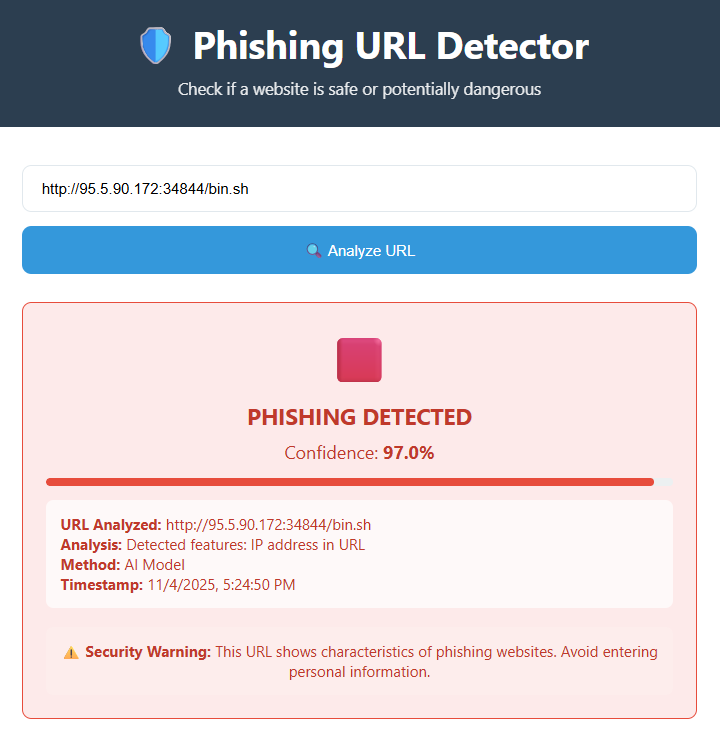}
    \centering
    \caption{Phishing URL Detector}
    \label{doc:phishing_url_detector}
\end{figure}

For better illustrating the phishing detection system, we develop a use case as shown in Fig.~\ref{doc:phishing_url_detector}, where 
a client enters an URL for the analysis in the dashboard.
When the URL is entered by the client, the \texttt{MCP Core} module sends the request to \texttt{Service Layer}, which analyzes the URL and provides the result as \texttt{phishing detected}.
Moreover, the \texttt{Phishing Detector} system also offers the rationale for the analysis.
As we can see in Fig.~\ref{doc:phishing_url_detector}, the system detects features such as the IP address in the URL, which signals it as a phishing.

\subsection{MCP Components}

Moreover, the MCP Server is designed to provide a secure, scalable, and explainable phishing detection framework by integrating \textit{Context Isolation}, \textit{Provenance Validation}, and \textit{Hybrid Feature Fusion}. These components collectively enhance robustness, interpretability, and generalization across heterogeneous phishing datasets.

\subsubsection{Context Isolation}

Context Isolation ensures that each phishing detection request is processed independently to prevent contextual interference and adversarial contamination. 
Let an incoming URL request be denoted as $r_i$, that is assigned to an isolated context $C_i$ defined as:

\begin{equation}
\label{ci}
C_i = \{ \mathbf{x}_i, f(\mathbf{x}_i), \hat{y}_i \}
\end{equation}

where $\mathbf{x}_i \in \mathbb{R}^n$ denotes the extracted feature vector, $f(\cdot)$ represents the trained machine learning classifier, and $\hat{y}_i \in \{0,1\}$ indicates the predicted class label (legitimate or phishing). 
Context isolation enforces:

\begin{equation}
\label{ci_cj}
C_i \cap C_j = \emptyset \quad \forall \; i \neq j
\end{equation}

This constraint ensures secure, independent inference and mitigates session-based and cross-request attacks.

\subsubsection{Provenance Validation}

Provenance Validation verifies the origin and trustworthiness of each input sample.
Each input is associated with a provenance label $p_i$ such as:

\begin{equation}
\label{pi}
p_i \in \{ D_{UCI}, D_{OpenPhish}, D_{EvilGinx}, D_{GenAI} \}
\end{equation}

where $D_{UCI}$, $D_{OpenPhish}$, $D_{EvilGinx}$, and $D_{GenAI}$ represent public benchmark data, phishing-tool-generated data, and Generative AI–generated data, respectively.
A \textit{Provenance Confidence Score (PCS)} is computed as:

\begin{equation}
\label{pcs}
PCS_i = \frac{1}{k} \sum_{j=1}^{k} \mathbb{I}(p_i = p_j)
\end{equation}

where $\mathbb{I}(\cdot)$ is the indicator function and $k$ indicates the number of nearest neighbors in the feature space.
Samples with low $PCS_i$ values are flagged for further examination to ensure data integrity.

\subsubsection{Hybrid Feature Fusion}

Hybrid Feature Fusion combines statistically significant features with explainability-driven insights.
Let $F_{IG}$ denote the feature subset selected using Information Gain, and $F_{XAI}$ represent the features identified using SHAP and LIME.
The final feature set is defined as:

\begin{equation}
\label{f_final}
F_{\text{final}} = F_{IG} \cup F_{XAI}
\end{equation}

Each feature $f_j \in F_{\text{final}}$ is assigned a composite importance weight:

\begin{equation}
\label{wj}
w_j = \alpha \cdot IG_j + \beta \cdot |f_j|, \quad \text{where } \alpha + \beta = 1
\end{equation}

This weighted fusion balances global statistical relevance with local interpretability.

\subsubsection{Decision Function}

The final phishing classification decision within the proposed MCP framework is achieved through a context-aware probabilistic inference. 
Given an incoming request $r_i$ with feature representation $\mathbf{x}_i$ and hybrid feature weights $\mathbf{w}$ derived from Information Gain and XAI-based importance scores, the classifier produces a weighted feature embedding $\mathbf{x}_i \odot \mathbf{w}$. 
The trained model $f(\cdot)$ then estimates the posterior class probabilities conditioned on the isolated context. 
Accordingly, the predicted label $\hat{y}_i$ is computed as:
\begin{equation}
\label{hat_y}
\hat{y}_i = \arg\max_{c \in \{0,1\}} 
P\bigl(c \mid f(\mathbf{x}_i \odot \mathbf{w})\bigr)
\end{equation}
where $c=0$ and $c=1$ correspond to legitimate and phishing classes, respectively. 
This formulation ensures that the final decision jointly reflects statistical relevance, explainability-driven feature importance, and strict context isolation enforced by the MCP server.

\subsection{MCP Evaluation Metrics}

We propose four evaluation metrics to analyze heterogeneous phishing datasets, the Context Integrity Score (CIS), Attack Propagation Factor (APF), Mitigation Response Efficiency (MRE), and Context Sensitivity Index (CSI). 

\subsubsection{Context Integrity Score (CIS)}
It measures contextual similarity between pre- and post-attack states, which derives from Eq.~\ref{ci}. It quantifies the similarity between the original isolated context  $C_i^{pre}$ and the attacked context $C_i^{post}$, which can be derived from Eq.~\ref{ci} as follows.

\begin{equation}
    CIS_i = sim(C_i^{pre}, \ C_i^{post})
\end{equation}

\subsubsection{Attack Propagation Factor (APF)}  It quantifies the degree of contextual degradation caused by adversarial manipulation, i.e., a direct consequence of violating isolation (Eq.~\ref{ci_cj}).
APF measures how much this constraint is broken under attack using Eq.~\ref{apf}.

\begin{equation}
\label{apf}
APF = \frac{1}{N} \sum_{i \neq j} \frac{C_i^{pre} \cap C_j^{post}}{|C_i|}
\end{equation}

\subsubsection{Context Sensitivity Index (CSI)}  It captures the susceptibility of model performance to small perturbations in contextual inputs, and would be linked to $F_{final}$ of Eq.~\ref{f_final}.
 CSI measures that how sensitive the prediction $\hat{y}$ is to small perturbations in $F_{final}$ and would be related as follows.

\begin{equation}
\label{csi}
CSI = \mathbb{E} [|P(\hat{y}|F_{final}) - P(\hat{y}|F_{final} + \delta)|]
\end{equation}

\subsubsection{Mitigation Response Efficiency (MRE)}  It measures how quickly and effectively the system restores contextual stability after detecting low-provenance or adversarial samples, and would be stated as in Eq.~\ref{mre}. 
\begin{equation}
\label{mre}
MRE = \frac{CIS_{post_mitigation}  - CIS_{post_attack}}{CIS_{pre_attack}} 
\end{equation}

The defined MCP components are directly operationalized through the proposed evaluation metrics. 
Context Isolation is quantified using the Context Integrity Score (CIS) and Attack Propagation Factor (APF), 
which respectively measure the preservation of isolated inference states and the extent of cross-context leakage under adversarial manipulation. 
Provenance Validation and Hybrid Feature Fusion influence the Context Sensitivity Index (CSI) by stabilizing model predictions against small contextual perturbations, 
while Mitigation Response Efficiency (MRE) captures the system’s ability to restore contextual integrity following attack detection. 
Together, these metrics provide a holistic evaluation of MCP’s robustness, resilience, and explainability.

\section{Result Evaluation}
\label{evaluation}

\subsection{Simulation-Based Evaluation}



To quantitatively evaluate the effectiveness of the proposed phishing detection framework, a set of standard performance metrics is used, including Accuracy, Precision, Recall, F1-Score and AUC.   
These metrics assess classification accuracy, detection capability, and robustness across different phishing datasets: UCI, OpenPhish, EvilGinx, and GenAI, as mentioned in Table~\ref{tab:class_distribution}. 
The average performance metrics for classical models, ensemble models, neural network models and transformer models are stated in Table~\ref{tab:classical}, Table~\ref{tab:ensemble}, Table~\ref{tab:neural}, and Table~\ref{tab:transformer}, respectively. 
Table~\ref{tab:coefficient_23features} shows the average model coefficients in 23 common features.

\begin{table}[h]
\centering
\caption{Performance Comparison of Classical Models}
\label{tab:classical}
\begin{tabular}{lccccc}
\hline
Model & Accuracy & Precision & Recall & F1 Score & ROC AUC \\
\hline
\textbf{Logistic Regression} & \textbf{0.9244} & \textbf{0.9315} & \textbf{0.9244} & \textbf{0.9243} & \textbf{0.9604} \\
SGD Classifier      & 0.9184 & 0.9286 & 0.9184 & 0.9180 & 0.9581 \\
Ridge Classifier    & 0.9172 & 0.9288 & 0.9172 & 0.9166 & 0.9594 \\
Elastic Net         & 0.8914 & 0.8862 & 0.8914 & 0.8825 & 0.9587 \\
\hline
\end{tabular}
\end{table}

\begin{table}[h]
\centering
\caption{Performance Comparison of Ensemble Models}
\label{tab:ensemble}
\begin{tabular}{lccccc}
\hline
Model & Accuracy & Precision & Recall & F1 Score & ROC AUC \\
\hline
\textbf{CatBoost}       & \textbf{0.950175} & \textbf{0.95295} & \textbf{0.950175} & \textbf{0.950225} & \textbf{0.978975} \\
XGBoost        & 0.949975 & 0.95235 & 0.949975 & 0.950025 & 0.979575 \\
Random Forest  & 0.948800 & 0.95120 & 0.948800 & 0.948875 & 0.976850 \\
Decision Tree  & 0.941550 & 0.94370 & 0.941550 & 0.941575 & 0.957750 \\
\hline
\end{tabular}
\end{table}

\begin{table}[h]
\centering
\caption{Average Performance Comparison of  Neural Network Models}
\label{tab:neural}
\begin{tabular}{lccccc}
\hline
Model & Accuracy & Precision & Recall & F1 Score & ROC AUC \\
\hline
MLP       & 0.939025 & 0.941075 & 0.939025 & 0.939025 & 0.966475 \\
\textbf{CNN}       & \textbf{0.940225} & \textbf{0.942325} & \textbf{0.940225} & \textbf{0.940300} & \textbf{0.968125} \\
GRU       & 0.931725 & 0.940200 & 0.931725 & 0.931650 & 0.962850 \\
CNN-LSTM  & 0.937900 & 0.940075 & 0.937900 & 0.937800 & 0.966800 \\
\hline
\end{tabular}

\end{table}



\begin{table}[h]
\centering
\caption{Average Performance Comparison of Transformer Models}
\label{tab:transformer}
\begin{tabular}{|l|c|c|c|c|c|}
\hline
Model & Accuracy & Precision & Recall & F1 Score & ROC AUC \\
\hline
BERT-PhishFinder & 0.9962 & 0.9993 & 0.9939 & 0.9966 & 0.9996 \\
CodeBERT         & 0.9955 & 0.9985 & 0.9934 & 0.9959 & 0.9995 \\
\textbf{DistilBERT}       & \textbf{0.9978} & \textbf{0.9995} & \textbf{0.9963} & \textbf{0.9979} & \textbf{1.0000} \\
BERT             & 0.9973 & 0.9995 & 0.9952 & 0.9973 & 0.9999 \\
\hline
\end{tabular}
\end{table}

\begin{table*}[h]
\centering
\caption{Average Model Coefficients across Datasets of 23 Common Features}
\label{tab:coefficient_23features}
\begin{tabular}{|l|c|c|c|c|}
\hline
\textbf{Feature} & \textbf{UCI} & \textbf{OPEN} & \textbf{EVIL } & \textbf{DS} \\
\hline
Abnormal\_URL & -0.4827 & 0.6688 & 1.6715 & 0.3482 \\
DNSRecord & 0.2437 & -2.1974 & -1.0089 & -2.3997 \\
Google\_Index & 0.6323 & -4.1267 & -1.0089 & -3.7476 \\
HTTPS\_token & -0.3112 & 0.0002 & -1.0089 & 0.0000 \\
Iframe & -0.2753 & 0.0000 & -1.0089 & 0.3228 \\
Links\_in\_tags & 0.8051 & 0.0000 & -1.0089 & 0.8991 \\
Links\_pointing\_to\_page & 0.4836 & -2.2492 & -1.0089 & -2.5735 \\
Prefix\_Suffix & 2.6478 & 1.8077 & 0.3514 & 0.4961 \\
Redirect & -1.1182 & 0.5051 & -1.0089 & -0.0085 \\
Request\_URL & 0.3498 & 0.0000 & 1.6715 & 0.0000 \\
RightClick & 0.2697 & 0.0000 & -1.0089 & 0.0000 \\
SFH & 0.9813 & 0.0223 & 1.6715 & 0.0000 \\
Shortening\_Service & -0.8048 & -0.4122 & -0.5280 & 0.0507 \\
Statistical\_report & 0.4847 & -2.3826 & -0.2661 & -3.1394 \\
Submitting\_to\_email & -0.2062 & 0.0000 & -1.0089 & 0.0000 \\
URL\_Length & -0.0635 & 0.0340 & 0.2053 & 0.0976 \\
URL\_of\_Anchor & 3.1479 & 0.0000 & -1.0089 & 0.4450 \\
double\_slash\_redirecting & 0.4681 & 0.0000 & 0.4209 & 0.3237 \\
having\_At\_Symbol & 0.4063 & 0.0045 & 0.2679 & 0.8544 \\
having\_IP\_Address & 0.4866 & 0.7305 & -0.2405 & 2.2674 \\
having\_Sub\_Domain & 0.6981 & 3.2913 & 2.5123 & -0.4731 \\
on\_mouseover & 0.0033 & 0.0000 & -1.0089 & 0.0000 \\
web\_traffic & 0.5755 & -2.3203 & -1.0089 & -3.4701 \\
\hline
\end{tabular}
\end{table*}

\begin{figure}[h!]
    \centering
    \begin{tikzpicture}
        \begin{axis}[
            ybar,
            bar width=9pt,
            width=0.7\textwidth,
            height=0.50\textwidth,
            ylabel={Score},
            xlabel={Dataset},
            ymin=0.9, ymax=1.01,
            xtick=data,
            symbolic x coords={UCI, OpenPhish, EvilGinx, GenAI},
            enlarge x limits=0.2,
            legend style={at={(0.5,-0.25)}, anchor=north, legend columns=-1},
            nodes near coords,
            every node near coord/.append style={
                font=\tiny,
                rotate=90,
                anchor=west,
                /pgf/number format/.cd,fixed,precision=3
            },
            grid=major,
        ]

        \addplot[fill=blue!70] coordinates {
            (UCI,0.955719) (OpenPhish,0.996482) (EvilGinx,0.979578) (GenAI,0.999823)
        };

        \addplot[fill=red!70] coordinates {
            (UCI,0.949993) (OpenPhish,0.998406) (EvilGinx,0.975253) (GenAI,1.000000)
        };

        \addplot[fill=green!70] coordinates {
            (UCI,0.959011) (OpenPhish,0.994692) (EvilGinx,0.979602) (GenAI,0.999687)
        };

        \addplot[fill=orange!70] coordinates {
            (UCI,0.954462) (OpenPhish,0.996541) (EvilGinx,0.977417) (GenAI,0.999844)
        };

        \addplot[fill=purple!70] coordinates {
            (UCI,0.993698) (OpenPhish,0.999928) (EvilGinx,0.997573) (GenAI,1.000000)
        };

        \legend{Accuracy, Precision, Recall, F1-Score, AUC}

        \end{axis}
    \end{tikzpicture}
    \caption{CatBoostClassifier performance across 4 phishing datasets}
    \label{fig:catboost_bar}
\end{figure}
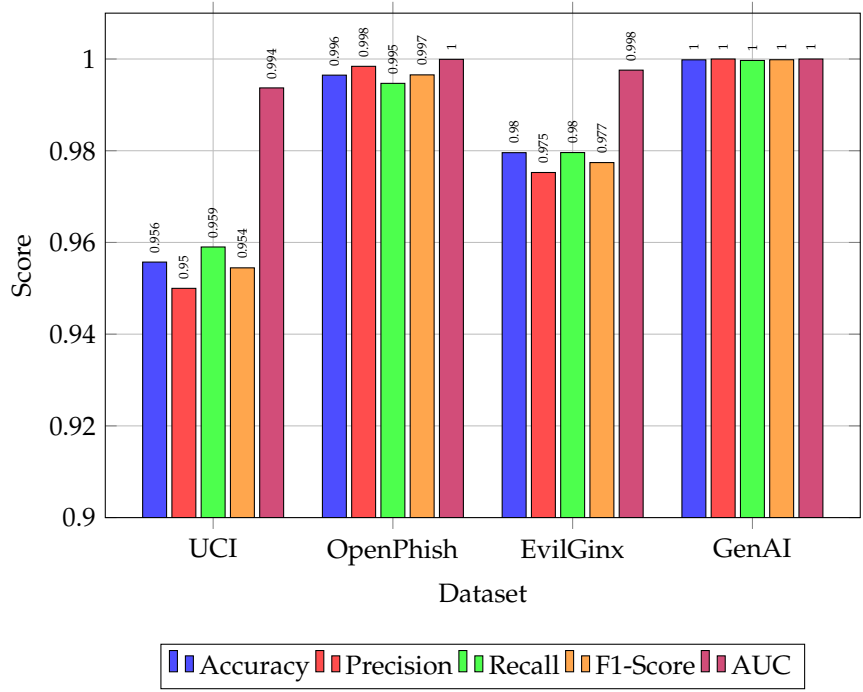

In the UCI dataset, Logistic Regression achieves 91.96\% accuracy, 91.18\% precision, 92.33\% recall, F1-score 91.75\%, and AUC 0.9757, while ElasticNet and SGDClassifier are slightly lower. On OpenPhish and GenAI, these linear models perform very well, reaching 99–100\% accuracy with AUC values close to 1.0. However, on EvilGinx, their performance drops (Logistic Regression 92.48\%, ElasticNet 92.66\%, SGDClassifier 91.94\%), indicating that simple linear models cannot fully capture complex phishing patterns.
Moreover,
Tree-based and boosting models consistently outperform linear models across all datasets. On UCI, Random Forest achieves 94.87\% accuracy (AUC 0.9887), XGBoost 95.21\% (AUC 0.9930), and CatBoost 95.57\% (AUC 0.9937). On OpenPhish, CatBoost performs best with 99.65\% accuracy and AUC 0.9999, while XGBoost and Random Forest are slightly lower. On EvilGinx, Random Forest and CatBoost maintain high accuracy (98.16\% and 97.95\%) and AUC above 0.997. Neural networks such as MLP, CNN, and GRU also perform well. On UCI, MLP achieves 94.28\% accuracy (AUC 0.9885), CNN 93.45\% (AUC 0.9858), and GRU 91.52\% (AUC 0.9733). On OpenPhish and GenAI, neural networks reach nearly and equal to 100\% accuracy and AUC. Figure~\ref{fig:catboost_bar} shows the overall accuracy comparison of different models on the hetrogeneous datasets.

\begin{figure}[htbp]
\centering
\begin{subfigure}[b]{0.45\textwidth}
\centering
\includegraphics[width=\textwidth]{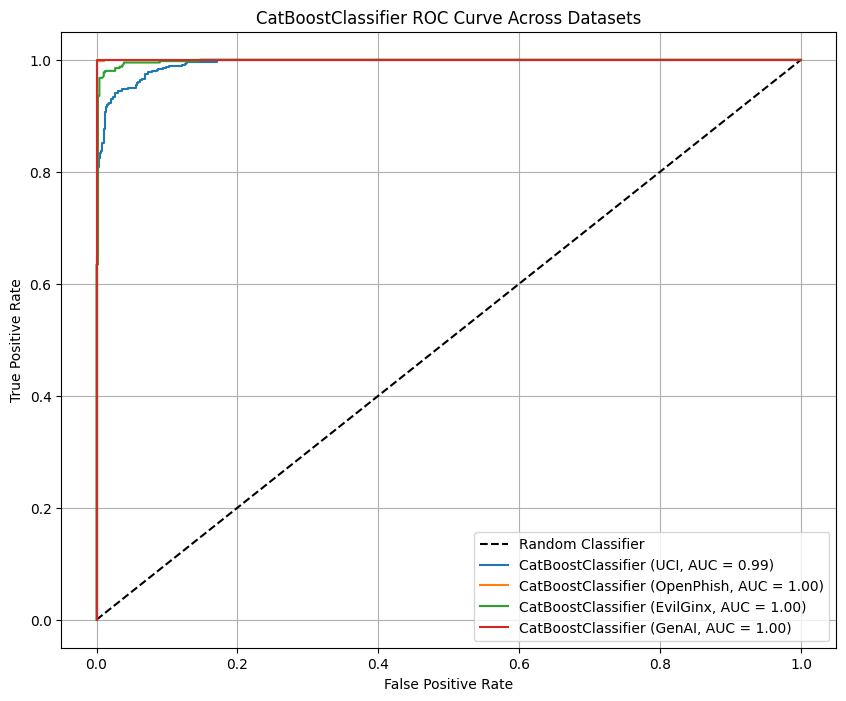}
\caption{ROC of four dataset}
\label{fig:roc}
\end{subfigure}
\hfill
\begin{subfigure}[b]{0.45\textwidth}
\centering
\includegraphics[width=\textwidth]{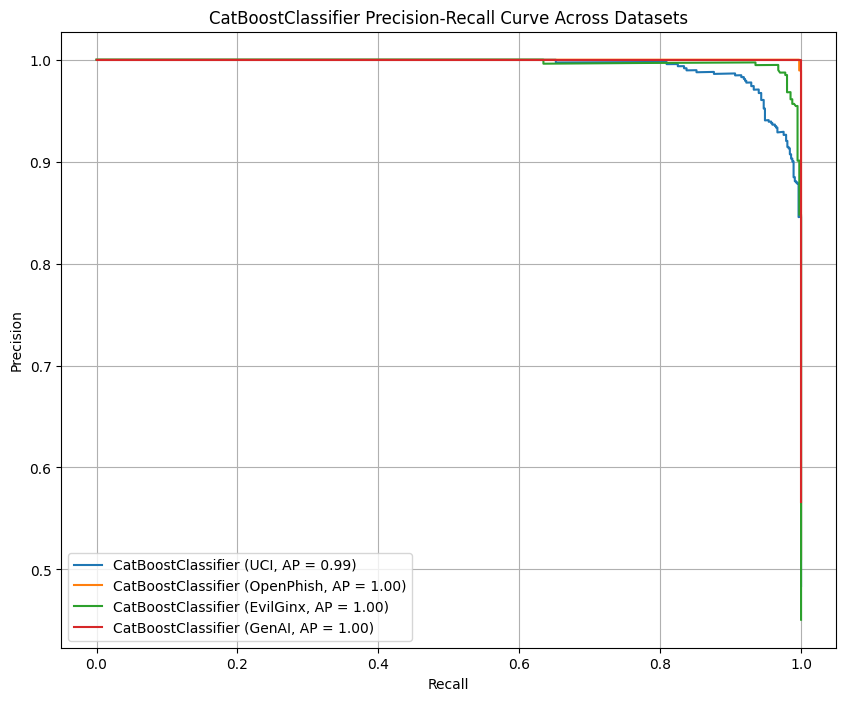}
\caption{Pre recall of four dataset}
\label{fig:recall}
\end{subfigure}
\end{figure}

Across all evaluation metrics, CatBoostClassifier consistently demonstrates superior and stable performance across heterogeneous phishing datasets, while CNN exhibits higher variance, particularly on the EvilGinx dataset, as summarized in Table~\ref{tab:catboost-performance}. Moreover, Figure~\ref{fig:roc} and Figure~\ref{fig:recall} confirms the best model claim against CatBoostClassifier.

\begin{table}[ht]
\centering
\caption{Performance of CatBoost across different phishing datasets}
\label{tab:catboost-performance}
\small
\begin{tabular}{|l|c|c|c|c|c|}
\hline
\textbf{Dataset} & \textbf{Accuracy} & \textbf{Precision} & \textbf{Recall} & \textbf{F1-Score} & \textbf{AUC} \\
\hline
UCI             & 0.9557 & 0.9499 & 0.9590 & 0.9545 & 0.9937 \\
\hline
OpenPhish       & 0.9965 & 0.9984 & 0.9947 & 0.9965 & 0.9999 \\
\hline
EvilGinx        & 0.9796 & 0.9753 & 0.9796 & 0.9774 & 0.9976 \\
\hline
GenAI           & 0.9998 & 1.0000 & 0.9997 & 0.9998 & 1.0000 \\
\hline
\end{tabular}
\end{table}




\subsection{Explainable AI (XAI) Based Evaluation}

\subsubsection{Information Gain}
Information Gain (IG) measures the dependency between input features and the target label. Higher IG values indicate that features are more useful for distinguishing phishing from legitimate samples.
The OpenPhish (0.136) and GenAI (0.123) datasets exhibit the highest average information gain, followed by EvilGinx (0.088). The UCI dataset shows the lowest information gain (0.035). These results confirm that features in OpenPhish and GenAI are more strongly aligned with phishing behavior, explaining their higher effectiveness in phishing detection tasks.

\subsubsection{SHAP Feature Importance Analysis}
We used SHAP values with the best CatBoost model to evaluate the impact of each feature on the model's decisions for detecting phishing URLs. On the UCI dataset, SHAP values range from approximately $\pm \mathbf{0.35}$ to $\pm \mathbf{0.40}$, indicating that a few key features have a very strong effect , the model relies heavily on these features to distinguish between legitimate and phishing URLs. For OpenPhish, the SHAP value range is smaller, around $\pm \mathbf{0.20}$, and EvilGinx is similar at $\pm \mathbf{0.22}$. This suggests that while features in these datasets are influential, their impact is more evenly distributed and not dominated by just one or two features. In contrast, the GenAI dataset exhibits almost zero SHAP values, only about $\pm \mathbf{0.003}$. This implies that no single feature significantly controls the decision; instead, the model achieves perfect results because legitimate and phishing URLs are naturally well separated in the dataset . For which GenAI dataset  accuracy touch to 100\% accuracy.

\begin{figure}[htbp]
\centering
\begin{subfigure}[b]{0.45\textwidth}
\centering
\includegraphics[width=\textwidth]{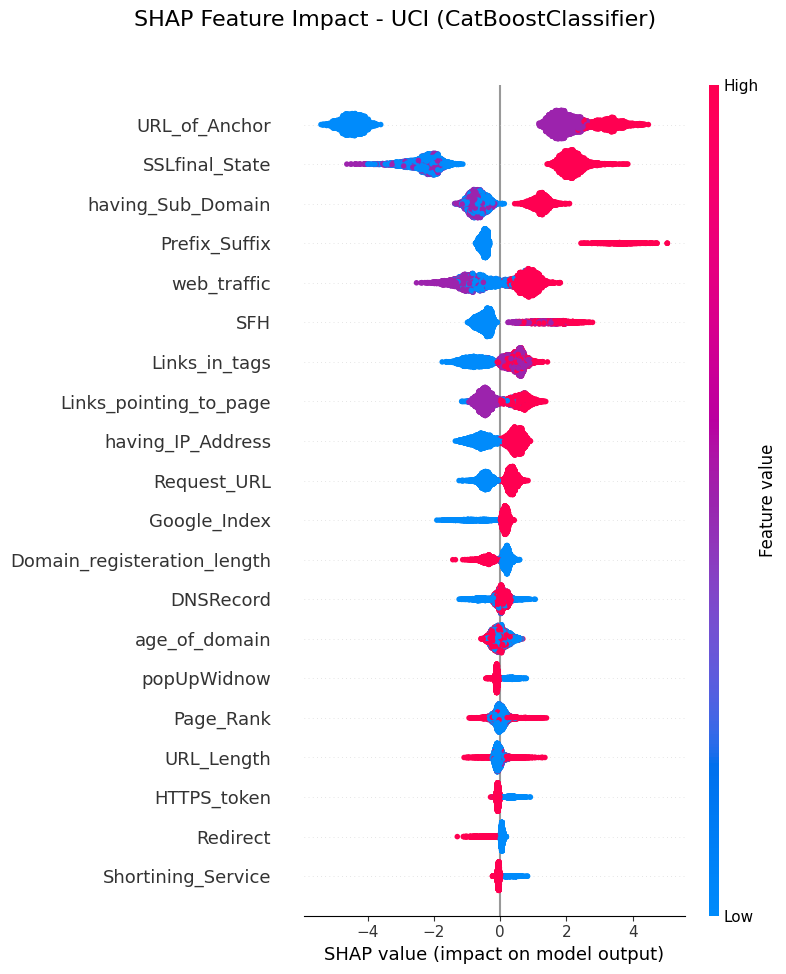}
\caption{SHAP for UCI Dataset}
\label{fig:1}
\end{subfigure}
\hfill
\begin{subfigure}[b]{0.45\textwidth}
\centering
\includegraphics[width=\textwidth]{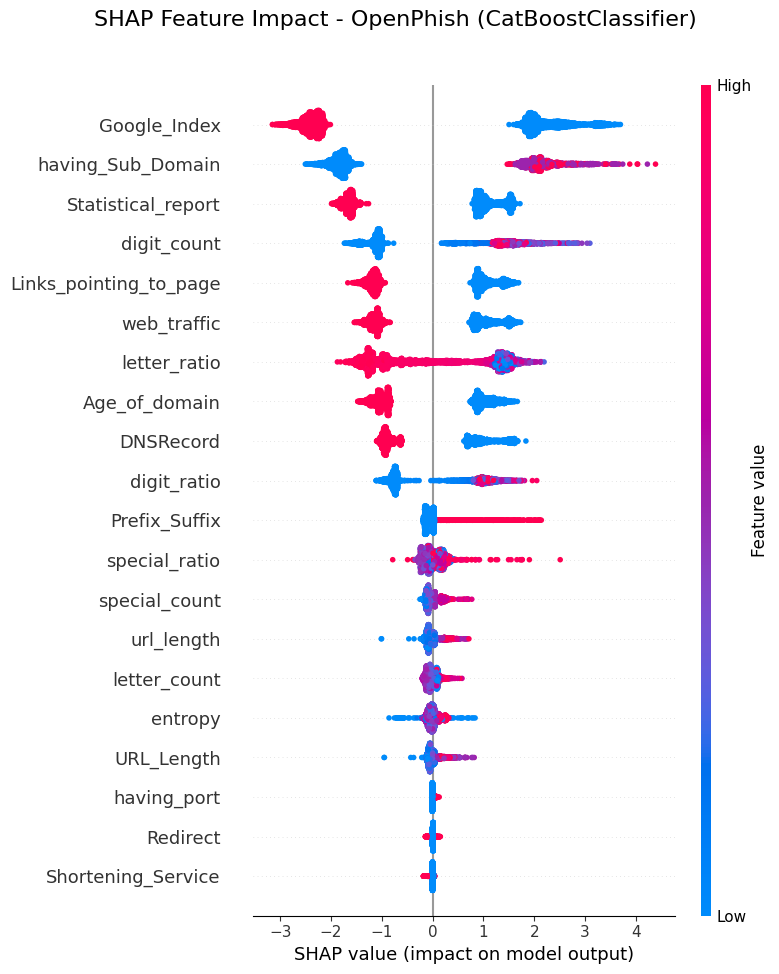}
\caption{SHAP for OpenPhish Dataset}
\label{fig:2}
\end{subfigure}

\vspace{1em} 

\begin{subfigure}[b]{0.45\textwidth}
\centering
\includegraphics[width=\textwidth]{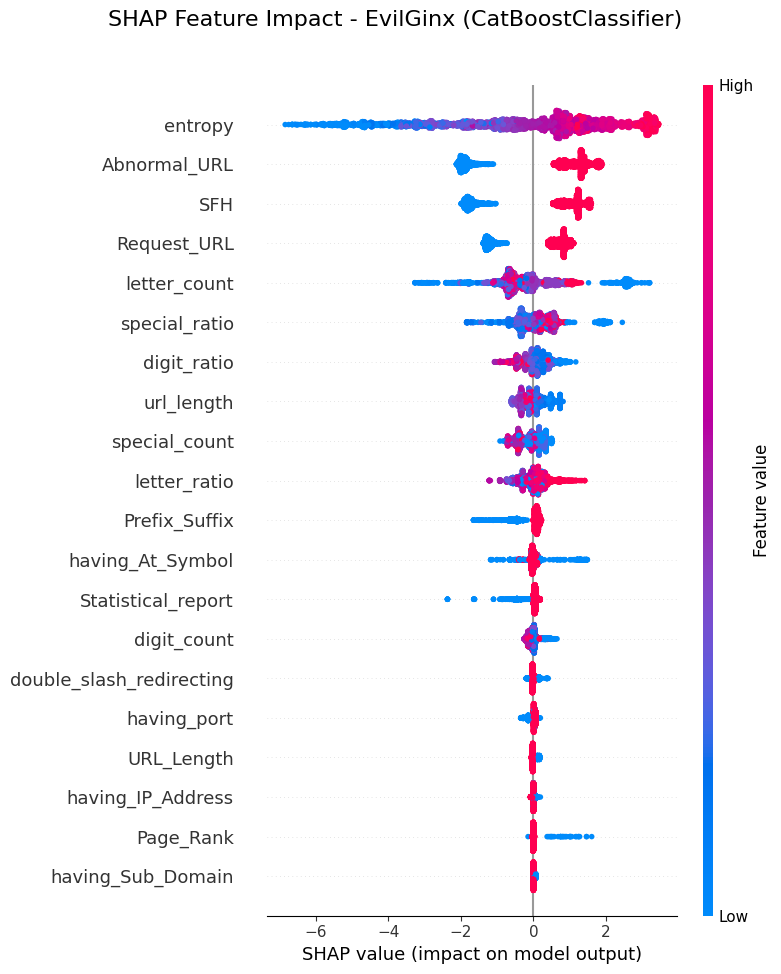}
\caption{SHAP for EvilGinx Dataset}
\label{fig:3}
\end{subfigure}
\hfill
\begin{subfigure}[b]{0.45\textwidth}
\centering
\includegraphics[width=\textwidth]{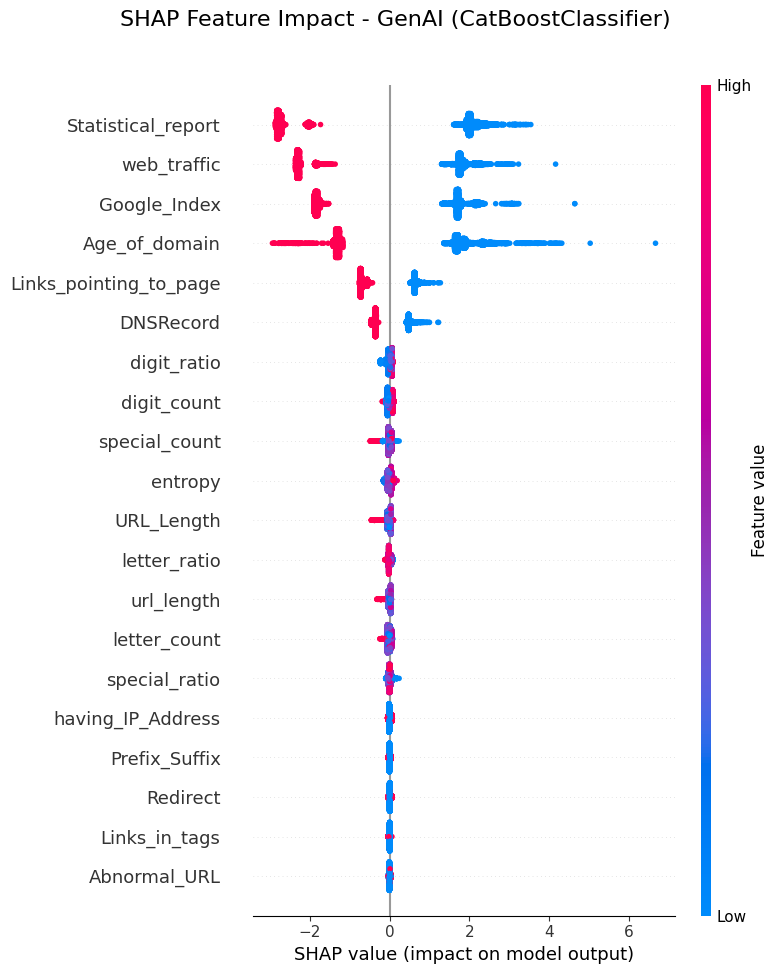}
\caption{SHAP for GenAI Dataset}
\label{fig:4}
\end{subfigure}
\caption{SHAP Analysis on Heterogeneous Dataset}
\label{fig:combined}
\end{figure}

\subsubsection{LIME Feature Importance Analysis}
LIME explanations provide local insights into individual predictions using CatBoostClassifier in the four datasets. It shows that which features are most critical for a single instance's classification.

LIME works by slightly perturbing the feature values of a single URL and observing how the model’s output changes. Based on these changes, it identifies the most influential features responsible for that specific prediction. In the LIME plots, green bars indicate features that increase the phishing score, while red bars indicate features that reduce the phishing score. The length of each bar represents the strength of the feature’s influence, and the text shown on each bar (for example, \texttt{SSLfinal\_State $\leq$ 1.00}) represents the exact condition applied by the model.

For the UCI dataset, the phishing classification is mainly driven by \texttt{SSLfinal\_State $\leq$ 1.00} ($\sim$0.38), \texttt{Prefix\_Suffix $\leq$ 1.00} ($\sim$0.30), \texttt{web\_traffic $\leq$ 1.00} ($\sim$0.22), and \texttt{having\_Sub\_Domain $\leq$ 1.00} ($\sim$0.18). These strong phishing indicators clearly outweigh the small benign contribution from \texttt{Links\_pointing\_to\_page $\leq$ 0.00} ($-0.07$). In the OpenPhish dataset, the model primarily relies on reputation-based and lexical features such as \texttt{Google\_Index = 0} ($\sim$0.31), \texttt{having\_Sub\_Domain} ($\sim$0.22), \texttt{Statistical\_report} ($\sim$0.20), and \texttt{Digit\_Features} ($\sim$0.18). For the EvilGinx dataset, phishing detection is dominated by \texttt{Abnormal\_URL} ($\sim$0.26), \texttt{SFH} ($\sim$0.23), and \texttt{entropy} ($\sim$0.20), even though \texttt{Digit\_Features} contributes negatively ($-0.11$), indicating conflicting feature behavior. In the GenAI dataset, the model depends mainly on \texttt{SSLfinal\_State} ($\sim$0.37), \texttt{Prefix\_Suffix} ($\sim$0.29), and \texttt{web\_traffic} ($\sim$0.21), while \texttt{Links\_pointing\_to\_page} has a minor negative influence ($-0.08$). This demonstrates that GenAI-based phishing URLs are detected primarily using general behavioral and domain-level signals rather than obvious structural URL manipulations.

\subsection{MCP-Based Evaluation}

Table~\ref{tab:use_case} summarizes the effectiveness of the proposed MCP-based framework across the UCI, OpenPhish, EvilGinx, and GenAI datasets. All datasets achieve a Context Integrity Score (CIS) of 1.0 after mitigation, highlighting that the recovery mechanism restores contextual consistency following attack injection. 
The Attack Propagation Factor (APF) indicates a dataset-specific variation, with measurable propagation observed only in the UCI dataset, whereas OpenPhish, EvilGinx, and GenAI show no attack propagation. 
The positive Mitigation Response Efficiency (MRE) for the UCI dataset shows measurable recovery from an initially degraded context, whereas other datasets maintain stable contextual behavior throughout the evaluation.

Table~\ref{tab:performance_comparison} further compares isolation, provenance validation, and hybrid mitigation strategies. 
The hybrid strategy preserves high CIS and low APF values, demonstrating that combining context isolation with provenance-aware validation improves resilience against contextual perturbations.
Overall, the results highlight that the proposed MCP-based framework improves contextual stability, supports attack traceability, and enhances robustness for phishing detection across heterogeneous datasets.
These findings also align with the explainability results, where the GenAI dataset shows a strong reliance on general behavioral and domain-level indicators rather than only obvious URL-level manipulations.


\begin{figure}[htbp]
\centering
\begin{subfigure}[b]{0.40\textwidth}
\centering
\includegraphics[width=\textwidth]{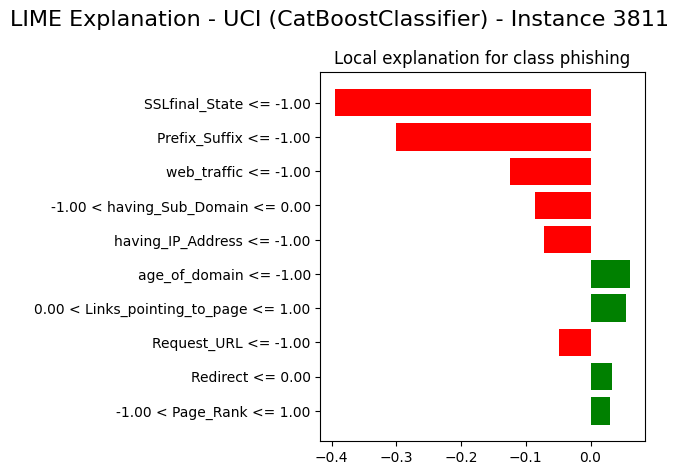}
\caption{LIME for UCI Dataset}
\label{LIME_UCI}
\end{subfigure}
\hfill
\begin{subfigure}[b]{0.45\textwidth}
\centering
\includegraphics[width=\textwidth]{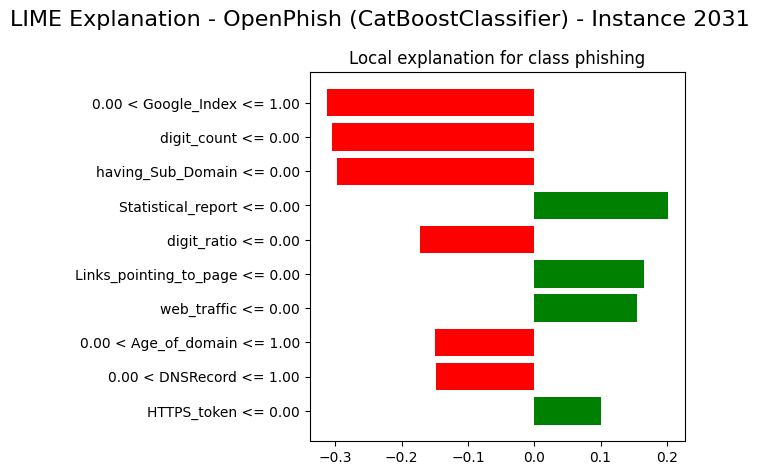}
\caption{LIME for OpenPhish Dataset}
\label{LIME_OpenPhish}
\end{subfigure}

\vspace{1em} 

\begin{subfigure}[b]{0.45\textwidth}
\centering
\includegraphics[width=\textwidth]{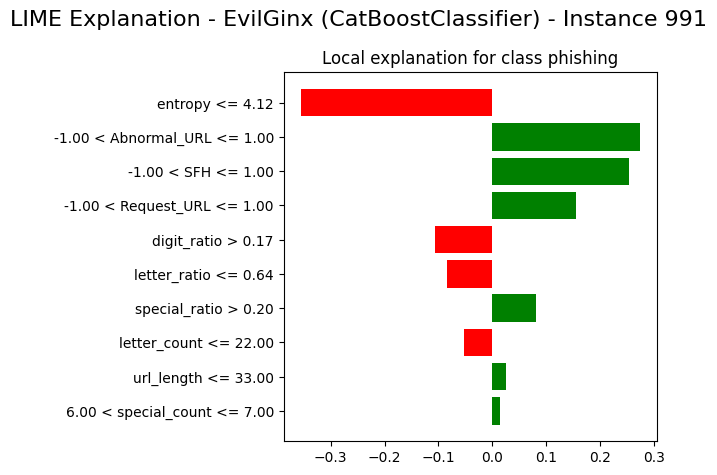}
\caption{LIME for EvilGinx Dataset}
\label{LIME_EvilGinx}
\end{subfigure}
\hfill
\begin{subfigure}[b]{0.45\textwidth}
\centering
\includegraphics[width=\textwidth]{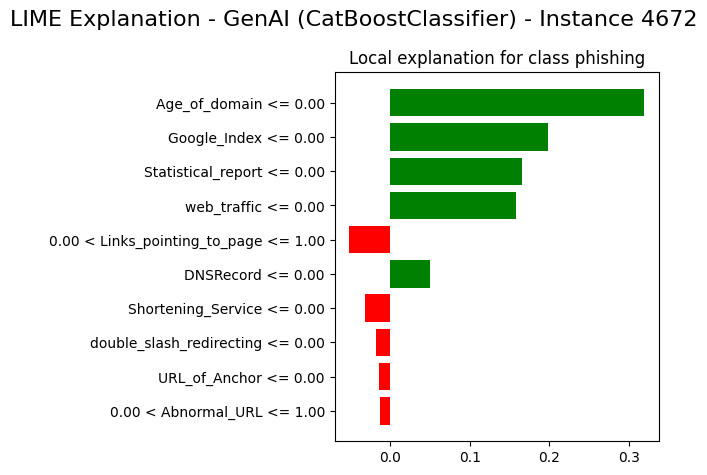}
\caption{LIME for GenAI Dataset}
\label{LIME_GenAI}
\end{subfigure}
\caption{LIME  Analysis on Heterogeneous Dataset}
\label{fig:LIME_All}
\end{figure}

\begin{table}[ht]
\centering
\caption{MCP-Based Attack–Mitigation Effectiveness Across Datasets}
\label{tab:use_case}
\begin{tabular}{|l|l|l|l|l|}
\hline
\textbf{Dataset} & \textbf{CIS} & \textbf{APF} & \textbf{MRE} & \textbf{CSI} \\
\hline
UCI & 1.0 & 0.368654 & 0.135505 &  0.99979\\
OpenPhish & 1.0 & 0.000000 & 0.000000 & 1.00000 \\			
EvilGinx & 1.0 & 0.000000 & 0.000000 &  1.00000\\
GenAI & 1.0 & 0.000000 & 0.000000 & 1.00000\\
\hline
\end{tabular}
\end{table}

\begin{table*}[htbp]
\centering
\caption{Performance Comparison of Isolation, Validation, and Hybrid Strategies Across Datasets}
\label{tab:performance_comparison}
\begin{tabular}{llcccc}
\hline
\textbf{Strategy} & \textbf{Dataset} & \textbf{CIS} & \textbf{APF} & \textbf{MRE} & \textbf{CSI} \\
\hline
\multirow{4}{*}{Isolation} 
 & UCI        & 0.8806 & 0.3668 & --      & 0.9998 \\
 & OpenPhish  & 1.0000 & 0.0000 & --      & 1.0000 \\
 & EvilGinx   & 1.0000 & 0.0000 & --      & 1.0000 \\
 & GenAI      & 1.0000 & 0.0000 & --      & 1.0000 \\
\hline
 \multirow{4}{*}{Validation}
 & UCI        & 1.0000 & 0.3646 & 0.1339 & 1.0000 \\
 & OpenPhish  & 1.0000 & 0.0000 & 0.0000 & 1.0000 \\
 & EvilGinx   & 1.0000 & 0.0000 & 0.0000 & 1.0000 \\
 & GenAI      & 1.0000 & 0.0000 & 0.0000 & 1.0000 \\
\hline
\multirow{4}{*}{Hybrid} 
 & UCI        & 1.0000 & 0.3663 & 0.1346 & 0.9999 \\
 & OpenPhish  & 1.0000 & 0.0000 & 0.0000 & 1.0000 \\
 & EvilGinx   & 1.0000 & 0.0000 & 0.0000 & 1.0000 \\
 & GenAI      & 1.0000 & 0.0000 & 0.0000 & 1.0000 \\
\hline
\end{tabular}
\end{table*}

The Mitigation Response Efficiency (MRE) is positive for the UCI dataset (0.1355), reflecting measurable recovery from an initially degraded context.  All datasets achieve CSI values close to 1.0, confirming that the overall contextual structure remains highly stable even under adversarial perturbations.  These results demonstrate that while mitigation mechanisms are effective across all datasets, GenAI-generated data exhibits the highest intrinsic robustness, showing no attack propagation and perfect context stability. This finding aligns with earlier entropy, information gain, and explainability analyses, and helps explain the superior machine learning performance observed on GenAI datasets.





\section{Conclusion, Limitations, and Future Work}
\label{conclusion}

This work presented an explainable phishing detection framework evaluated across heterogeneous datasets, to study diverse phishing patterns. 
Classical machine learning, ensemble learning, deep learning, and transformer-based models are compared, with ensemble and transformer-based models showing strong detection performance and DistilBERT achieving the best overall result among the evaluated transformer models.

The framework also includes Information Gain, SHAP, and LIME to improve explainability, along with an MCP-enabled phishing analysis system to facilitate practical deployment.
The MCP-based evaluation shows that context isolation, provenance-aware validation, and hybrid feature fusion improve contextual robustness and reduce attack propagation across heterogeneous datasets.

This work has some limitations. 
The evaluation mainly uses URL-based and derived numerical features, while real-world phishing detection may also require email content, webpage screenshots, HTML structure, and network-flow data.
In addition, tool-generated and AI-generated datasets are created in controlled environments and may not fully represent all live phishing behaviors.
Future work will therefore focus on multi-modal phishing detection, real-time deployment validation, adaptive context policies, and federated MCP-based detection for collaborative defense.

 \bibliographystyle{plain}
\small
\bibliography{references}

\end{document}